\def\itshape{\fontshape\itdefault\selectfont\let\mathrm=\mathit}

\documentclass[twoside,a4paper,12pt]{article}
\usepackage{a4p}
\usepackage{citesort}
\usepackage[final]{graphics}

\newcommand{\barMS}{\overline{\rm MS}}
\newcommand{\alphas}{\ensuremath{\mathrm{\alpha_s}}}

\newcommand{\rd}        {\ensuremath{\mathrm{d}}}

\newcommand{\be}{\begin{equation}}
\newcommand{\ee}{\end{equation}}
\newcommand{\bea}{\begin{eqnarray}}
\newcommand{\eea}{\end{eqnarray}}
\newcommand{\nn}{\nonumber}
\newcommand{\p}{\phantom}
\newcommand{\D}{\displaystyle}
\begin{document}
\title{\bf On the determination of 
  $\alphas$ from hadronic $\tau$ decays with
contour-improved, fixed order and renormalon-chain perturbation theory}
\author{Sven~MENKE\\%
        \normalsize\it Max-Planck-Institut f{\"u}r Physik\\%
        \normalsize\it (Werner-Heisenberg-Institut)\\%
        \normalsize\it F{\"o}hringer Ring 6, D-80805 M{\"u}nchen, Germany}

\maketitle

\begin{abstract} 
  \noindent One of the largest theoretical uncertainties assigned to
  the strong coupling constant \alphas\ as determined from hadronic
  tau decays stems from the differences in the results for Fixed Order
  Perturbation Theory (FOPT), Contour Improved Perturbation Theory
  (CIPT) and Renormalon Chain Perturbation Theory (RCPT). It is often
  argued that the three methods differ in the treatment of higher
  orders only and therefore the full difference should be treated as
  theoretical error. Recently other arguments either in favor of FOPT,
  CIPT or RCPT have been given, but none of those is able to combine
  all three to a single value in the strong coupling constant. In this
  note I will show that FOPT alone has a much larger uncertainty than
  previously assumed and therefore agrees within error with
  CIPT. Furthermore a more appropriate matching of the different
  schemes used in RCPT reduces the difference to the CIPT result by a
  factor of 6.  Together with recently published results for the 4th
  order term $K_4$ this reduces the theoretical error on \alphas\ by a
  factor of $2.5$ compared to the previously assumed spread of the
  three perturbative approaches.
\end{abstract}

\section{Introduction}
Hadronic decays of the $\tau$ lepton are among the most actively
studied fields in QCD. The unique situation of a small mass scale and
still small non-perturbative contributions allow for a very precise
determination of the strong coupling constant
$\alphas$~\cite{Braaten:1988hc,Braaten:1989ea,Braaten:1992qm,LeDiberder:1992fr}
(For recent reviews
see~\cite{Korner:2000xk,Milton:2000fi,Cvetic:2001sn,Cvetic:2001ws,Davier:2008sk,Beneke:2008ad}).
The ratio of the hadronic decay width of the $\tau$ and its leptonic
width can be written as
\begin{equation}
  \label{eq:r_tau}
  R_{\tau} = 3 S_{\rm EW} \left(\left|V_{\rm ud}\right|^2
    +\left|V_{\rm us}\right|^2\right)\,
  \left(1+\delta_{\rm EW}'+\delta_{\rm pert}+\delta_{\rm non-pert}\right),
\end{equation}
where $S_{\rm EW} =
1.0198\pm0.0006$~\cite{Marciano:1988vm,Davier:2002dy} and $\delta_{\rm
  EW}'= 0.0010\pm0.0010$~\cite{Braaten:1990ef} are small electroweak
corrections, $\delta_{\rm non-pert}$ denotes a $O($few$\%)$
non-perturbative correction and $\delta_{\rm pert}$ is the
perturbative prediction.  Neglecting the masses of the quarks (as is a
good approximation for the non-strange decay width of the $\tau$) the
perturbative part is given by
\begin{equation}
  \label{eq:delta_pert}
  1+\delta_{\rm pert} = \sum_{n=0}^4{\big.
    \frac{K_n}{2\pi i}\!\!\!\!\oint\limits_{|s|=m_\tau^2}{\!\!\!\! 
      \frac{\rd s}{s} \left(
        1 - 2 \frac{s}{m_\tau^2} + 2 \frac{s^3}{m_\tau^6} - \frac{s^4}{m_\tau^8}
      \right) \left(\frac{\alphas(-s)}{\pi}\right)^n\!\!}} + O(\alphas^5),
\end{equation}
with the known coefficients~\cite{Chetyrkin:1979bj,Dine:1979qh,Celmaster:1980xr,Gorishnii:1991vf,Surguladze:1991tg,Baikov:2008jh}
\begin{eqnarray}\label{eq:kn}
  K_0  & = &  K_1  =  1, \nn\\
  K_2  & = & \frac{299}{24} - 9\,\zeta(3), \nn\\ 
  K_3^{\overline{\rm MS}} &  = & \frac{58057}{288} -
  \frac{779}{4}\zeta(3) + \frac{75}{2}\zeta(5), \\
  K_4^{\overline{\rm MS}} & = & \frac{78631453}{20736}+
  \frac{4185}{8}\zeta(3)^2 - \frac{1704247}{432}\zeta(3) + 
  \frac{34165}{96}\zeta(5) - \frac{1995}{16}\zeta(7)  \nn,
\end{eqnarray}
where $K_3$ and $K_4$ are scheme dependent and given here in the
$\overline{\rm MS}$-scheme and $\zeta(n)$ denotes the Riemann zeta
function. The 4-th order term $K_4^{\overline{\rm MS}}\simeq49.0757$
deviates substantially from previous estimates and partial
calculations of that coefficient $K_4^{\rm partial} =
27\pm16$~\cite{Kataev:1995vh,Baikov:2002uw}. The fifth-order term has
been estimated to $K_5 \simeq 275$ in~\cite{Baikov:2008jh}, but the
large deviation of the exact $K_4$ from it's prediction suggests that
a $100\%$ error on $K_5$ is realistic.  For the purpose of evaluating
differences stemming from the 5th and higher orders I'll use $K_5 =
400\pm400$ in this note.

The methods FOPT and CIPT~\cite{LeDiberder:1992fr} differ in the way
(\ref{eq:delta_pert}) is calculated.  In the CIPT approach the
$\beta$-function is used to get numerical solutions for $\alphas(-s)$
in the complex $s$-plane by starting with $\alphas(m_\tau^2)$.  The
integrand is thus calculated in small steps on the circle
$|s|=m_\tau^2$ and the sum of all pieces gives the total integral.

For the FOPT method the $\beta$-function and its derivatives are
Taylor expanded in $s$ around $s_0=m_\tau^2$ which leads to a power
series representation of $\alphas(-s)$ in powers of
$\alphas(m_\tau^2)$.  The series is truncated at the desired order
(here the 5th) in the strong coupling and inserted in the integral
which becomes solveable now. The usual FOPT result reads:
\begin{eqnarray}
  \label{eq:fopt}
  & & \delta_{\rm pert} = 
  \frac{\alphas(m_\tau^2)}{\pi} + 5.2023
  \frac{\alphas^2(m_\tau^2)}{\pi^2} + 26.366 
  \frac{\alphas^3(m_\tau^2)}{\pi^3} + \\
  & & \qquad 127.08 \frac{\alphas^4(m_\tau^2)}{\pi^4} + (K_5 + 307.78)
  \frac{\alphas^5(m_\tau^2)}{\pi^5} + 
  O(\alphas^6).\nn
\end{eqnarray}
As is demonstrated in~\cite{Baikov:2002uw,Baikov:2008jh} the fourth
and fifth order terms contribute very little to the perturbative part
and the difference between the FOPT and the CIPT result is much larger
than the contributions from these terms even if generous errors are
used for $K_5$.  Taking $\alphas(m_\tau^2)=0.35$ and $K_5=400$ as
reference values we could first calculate $\delta_{\rm pert}$ from the
CIPT approach and extract $\alphas(m_\tau^2)$ again using FOPT:
\begin{equation}
  \label{eq:dpertcipt}
  \delta_{\rm pert}^{\rm CIPT}(\alphas(m_\tau^2)=0.35) = 0.21179, 
\end{equation}
\begin{equation}
  \label{eq:asfopt}
  \alphas(m_\tau^2)^{\rm FOPT}(\delta_{\rm pert}=0.21179) = 0.32543. 
\end{equation}
The deviation of either value from their mean is with $\Delta\alphas =
\pm 0.012$ almost twice as large as the uncertainty due to higher
orders $\Delta\alphas_{\Delta K_5} = 0.007$.  The reason for this
large difference is the choice of the point on the circle $|s| =
m_\tau^2$ in the complex $s$-plane around which the $\beta$-function
and its derivatives are Taylor expanded to approximate the strong
coupling on the circle. In the following section the FOPT formalism
will be generalized to allow for other choices.

\section{Generalized FOPT}\label{sec:genFOPT}
The starting point is the perturbative expansion of the $\beta$
function, which is given by
\begin{equation}
  \beta(a_{\rm s}) = \frac{{\rm d}a_{\rm s}(s)}{{\rm d}\ln s} = -\beta_0 a_{\rm s}^2(s)
  -\beta_1 a_{\rm s}^3(s) -\beta_2 a_{\rm s}^4(s) -\beta_3 a_{\rm
    s}^5(s) - \dots,\label{eq:beta}
\end{equation}
where $a_{\rm s}(s) = \alpha_{\rm s}(s)/(4\pi)$.  The first two terms
in the
$\beta$-function~\cite{Gross:1973id,Politzer:1973fx,Caswell:1974gg,Jones:1974mm,Egorian:1979zx}
for $n_{\rm f}$ quark flavors,
\begin{eqnarray*}
  \beta_0 & = & 11 - \frac{2}{3} n_{\rm f}, \\ \beta_1 & = & 102 -
  \frac{38}{3} n_{\rm f},
\end{eqnarray*}
are universal at leading twist whereas the higher order terms are
scheme dependent.  In the $\barMS$ scheme the first two scheme
dependent coefficients are
known~\cite{Tarasov:1980au,Larin:1993tp,vanRitbergen:1997va,Czakon:2004bu}:
\begin{eqnarray}
  \beta_2^{\overline{\rm MS}} & = & \frac{2857}{2} - \frac{5033}{18}
  n_{\rm f} + \frac{325}{54} n_{\rm f}^2, \nonumber\\
  \beta_3^{\overline{\rm MS}} & = & \frac{149753}{6} + 3564\,\zeta(3) -
  \left(\frac{1078361}{162} + \frac{6508}{27} \zeta(3)\right) n_{\rm f} +
  \nonumber\\ & & \qquad \left(\frac{50065}{162} + \frac{6472}{81}\zeta(3)
  \right) n_{\rm f}^2 + \frac{1093}{729} n_{\rm
    f}^3. \label{eq:beta_2}
\end{eqnarray}

The Taylor expansion of the evolution equation~(\ref{eq:beta}) around
$s_0$ reads up to the fifth order in \alphas:
\begin{eqnarray}
  \frac{\alphas(s)}{\pi} & = & 
  \frac{\alphas(s_0)}{\pi} - 
  \frac{1}{4}\beta_0\ln\frac{s}{s_0}
  \left(\frac{\alphas(s_0)}{\pi}\right)^2 + \nonumber\\
  & & \frac{1}{16}\left(\beta_0^2\ln^2\frac{s}{s_0}-\beta_1
    \ln\frac{s}{s_0}\right)\left(\frac{\alphas(s_0)}{\pi}\right)^3 -
  \nonumber\\
  & & \frac{1}{128}\left(2\,\beta_0^3\ln^3\frac{s}{s_0}-5\,\beta_0\beta_1
    \ln^2\frac{s}{s_0}+2\,\beta_2\ln\frac{s}{s_0}\right)
  \left(\frac{\alphas(s_0)}{\pi}\right)^4 + \nonumber\\
  & & \frac{1}{1536}\left(6\,\beta_0^4\ln^4\frac{s}{s_0}-
    26\,\beta_0^2\beta_1\ln^3\frac{s}{s_0}+\right.\nonumber\\
  & & \p{\frac{1}{1536}\biggl(}\left. 9\left(\beta_1^2+2\,\beta_0\beta_2\right)
    \ln^2\frac{s}{s_0}-6\,\beta_3\ln\frac{s}{s_0}\right)
  \left(\frac{\alphas(s_0)}{\pi}\right)^5 + O(\alphas(s_0)^6).
\label{eq:a(s)_taylor}
\end{eqnarray}
It should be noted that eq.~(\ref{eq:a(s)_taylor}) is strictly
speaking not a Taylor approximation since the truncation occurs at a
certain power of $\alphas$ and not at a certain power in the expansion
variable $\ln(s/s_0)$.
\begin{figure}[htb]
  \begin{center}
    \resizebox{0.49\textwidth}{!}{%
      \includegraphics{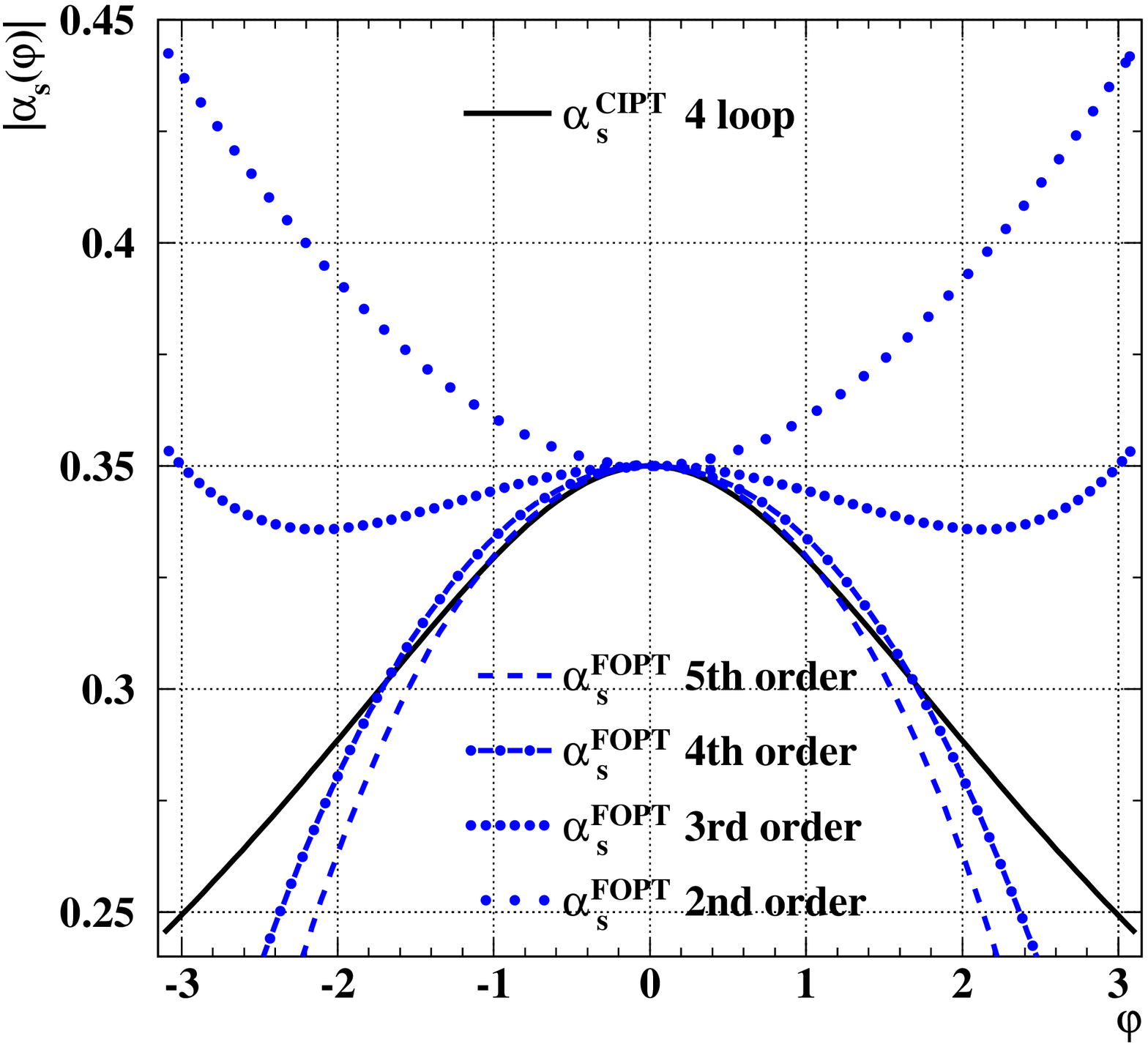}}
    \resizebox{0.49\textwidth}{!}{%
      \includegraphics{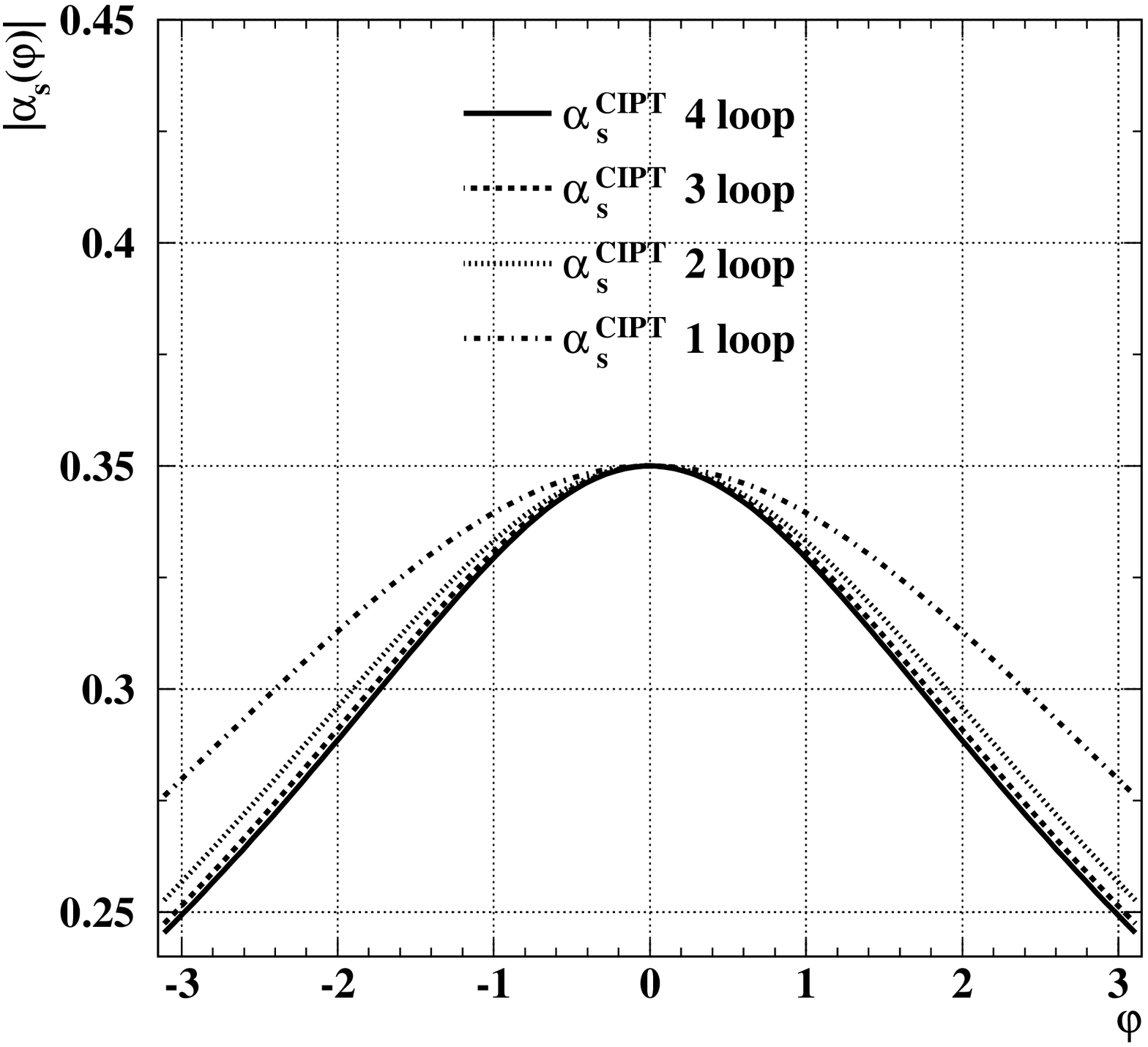}}
    \caption{\it Quality of the Taylor expansion of $\alphas(m_\tau^2
      \exp(i\varphi))$. The two plots show the absolute value
      $|\alphas|$ on the complex circle $s=m_\tau^2 \exp(i\varphi)$.
      The left plot shows with long-dashed, dash-dotted, narrow dotted
      and wide dotted lines the Taylor (FOPT) expansion up to 5th,
      4th, 3rd, and 2nd order, respectively. The 4-loop result for the
      numerically solved $\beta$-function (CIPT) is drawn as solid
      line for comparison.  The right plot shows with dahsed, dotted,
      and dash-dotted lines the numerical solutions for 3, 2 and 1
      loop $\beta$-functions, respectively.  Again the same 4-loop
      result as in the left plot is shown as a solid line.  The
      reference value $\alphas(m_\tau^2) = 0.35$ was used for all
      curves.}
    \label{fig:alphas0}
  \end{center}
\end{figure}

Since both the integrand in eq.~(\ref{eq:delta_pert}) and
eq.~(\ref{eq:a(s)_taylor}) are power series in $\alphas$ it is
interesting to compare the magnitudes of the coefficients in these
series. The largest values in eq.~(\ref{eq:a(s)_taylor}) are obtained
at $s=s_0 \exp(-i\pi)$ where the $|c_n|$ (the magnitude of the
coefficient in front of $(\alphas/\pi)^n$) read $1$, $7.07$, $51.52$,
$390.8$, $3023.85$.  At $s=s_0 \exp(-i\pi/2)$ (the average distance
from $s_0$) the $|c_n|$ are $1$, $3.53$, $13.98$, $62.33$, $275.16$.
These numbers grow much (slightly) faster at $-\pi$ ($-\pi/2$) than
the $K_n$, where (starting with $K_1$) we have $1$, $1.64$, $6.37$,
$49.08$, $\sim275$. Therefore it is conceivable that the nature of the
Taylor expansion of $\alphas$ dominates the uncertainty of the FOPT
result and not missing higher order $K_n$ terms.  To illustrate this
the Taylor expansion of $\alphas$ is modified in the following.

Figure~\ref{fig:alphas0} shows that the deviation of the Taylor
expanded \alphas\ from the numerically solved \alphas\ grows with the
distance of $s=s_0\exp(i\varphi)$ from the chosen development point
$s_0=m_\tau^2$. The CIPT results for 1 to 4-loop treatment on the
circle are also shown. Starting at 2-loop level the CIPT results are
almost indistinguishable while the FOPT deviations remain large even
at 5th order! Thus the correct treatment of the logarithms and not
higher orders in \alphas\ are the dominant source of uncertainty.

Furthermore the choice $s_0=m_\tau^2$ for FOPT is completely
arbitrary.  This becomes even more obvious in the light of the usual
procedure of comparing different \alphas\ measurements by evolving
them to the Z$^0$-mass with a numerically solved $\beta$-function.  If
we were to evolve the \alphas\ from tau-decays to the Z$^0$-mass in a
FOPT-like manner with equation~(\ref{eq:a(s)_taylor}) in just 3 steps
(with the quark flavor transitions to $n_f=4$ and $n_f=5$ at $m_\tau$
and $m_{\rm b}$, respectively) the numerical value of $\alphas(m_{\rm
  Z}^2)$ (for $\alphas(m_\tau^2) = 0.35$) would be $0.042$, $0.206$,
$0.064$, $0.160$, for the $1$, $2$, $3$, $4$ loop beta function,
respectively. Compared to the usual procedure of contour-improved
evolution of the coupling (i.e.~using eq.(\ref{eq:a(s)_taylor}) in
small steps) which yields $0.128$, $0.123$, $0.122$, $0.122$, for $1$,
$2$, $3$, $4$ loop, respectively, the 3 step FOPT solutions converge
very slowly, giving alternatingly lower and higher estimates of the
coupling as more and more orders in \alphas\ are considered and still
at 5th order resulting in a numerical value that is far below the
exact result. The FOPT terms up to the 5th order do not compensate for
the neglected large logarithms.

It is therefore natural to generalize this in the case of the $\tau$
and first evolve $\alphas(m_\tau^2)$ to
$\alphas(m_\tau^2\exp(i\varphi_0))$ with the numerically solved
$\beta$-function and derive the Taylor series of $\delta_{\rm pert}$
around this new point. The integral~(\ref{eq:delta_pert}) can in fact
be split in two pieces around $\varphi_0$ and $-\varphi_0$ since the
strong coupling at $-\varphi_0$ is just the complex conjugate of the
strong coupling at $\varphi_0$:
\begin{equation}
  \label{eq:ascplx}
  \alphas(m_\tau^2\exp(-i\varphi_0)) = \alphas(m_\tau^2\exp(i\varphi_0))^*
\end{equation}

The resulting $\delta_{\rm pert}$ up to the fifth order reads:
\begin{eqnarray}
  \label{eq:fopt_mod}
  \lefteqn{\delta_{\rm pert} = 
    a + \frac{8\, b}{3\, \pi} + a\, b\, \left(\frac{16\, {K_2}}{3\, \pi} - \frac{{\beta_0}\, {\varphi_0}}{2} + \left(\frac{16}{9\, \pi} - \frac{\pi}{4}\right)\, {\beta_0}\right)} \nn\\
  & & + \left(a^2 - b^2\right)\, \left({K_2} + \frac{19\, {\beta_0}}{48} + \frac{2\, {\beta_0}\, {\varphi_0}}{3\, \pi}\right) \nn \\
  & & + \left(a^3 - 3\, a\, b^2\right)\, \left( - \frac{{{\beta_0}}^2\, {{\varphi_0}}^2}{16} + \left(\frac{4}{9\, \pi} - \frac{\pi}{16}\right)\, {{\beta_0}}^2\, {\varphi_0} + \left(\frac{265}{1152} - \frac{{\pi}^2}{48}\right)\, {{\beta_0}}^2 \right.\nn\\
  & & \qquad \left. + \frac{4\, {K_2}\, {\beta_0}\, {\varphi_0}}{3\, \pi} + \frac{19\, {K_2}\, {\beta_0}}{24} + \frac{{\beta_1}\, {\varphi_0}}{6\, \pi} + {K_3} + \frac{19\, {\beta_1}}{192}\right) \nn\\
  & & - \left(b^3 - 3\, a^2\, b\right)\, \left( - \frac{{{\beta_0}}^2\, {{\varphi_0}}^2}{6\, \pi} - \frac{19\, {{\beta_0}}^2\, {\varphi_0}}{96} + \left(\frac{13}{27\, \pi} - \frac{19\, \pi}{192}\right)\, {{\beta_0}}^2 - \frac{{K_2}\, {\beta_0}\, {\varphi_0}}{2} \right.\nn\\
  & & \qquad \left. + \left(\frac{16\, {K_2}}{9\, \pi} - \frac{\pi\, {K_2}}{4}\right)\, {\beta_0} - \frac{{\beta_1}\, {\varphi_0}}{16} + \frac{8\, {K_3}}{3\, \pi} + \left(\frac{2}{9\, \pi} - \frac{\pi}{32}\right)\, {\beta_1}\right) \nn\\
  & &  + \left(a^4 - 6\, a^2\, b^2 + b^4\right)\, \left({K_4} + \frac{19\, {\beta_2}}{768} - \frac{19\, {{\beta_0}}^3\, {{\varphi_0}}^2}{256} + {{\beta_0}}^2\, \left(\frac{265\, {K_2}}{384} - \frac{{\pi}^2\, {K_2}}{16}\right)\right.\nn\\
  & & \qquad \left. - \left(\frac{19\, {\pi}^2}{768} - \frac{3355}{18432}\right)\, {{\beta_0}}^3 + \frac{19\, {K_2}\, {\beta_1}}{96} + \frac{19\, {K_3}\, {\beta_0}}{16} - \frac{3\, {K_2}\, {{\beta_0}}^2\, {{\varphi_0}}^2}{16} \right.\nn\\
  & & \qquad \left. - \left(\frac{5\, {\pi}^2}{384} - \frac{1325}{9216}\right)\, {\beta_0}\, {\beta_1} + \left(\frac{13}{36\, \pi} - \frac{19\, \pi}{256}\right)\, {{\beta_0}}^3\, {\varphi_0} \right.\nn\\
  & & \qquad \left. + {{\beta_0}}^2\, {\varphi_0}\, \left(\frac{4\, {K_2}}{3\, \pi} - \frac{3\, \pi\, {K_2}}{16}\right) - \frac{{{\beta_0}}^3\, {{\varphi_0}}^3}{24\, \pi} + \frac{{\beta_2}\, {\varphi_0}}{24\, \pi} - \frac{5\, {\beta_0}\, {\beta_1}\, {{\varphi_0}}^2}{128} \right.\nn\\
  & & \qquad \left. + \frac{{K_2}\, {\beta_1}\, {\varphi_0}}{3\, \pi} + \frac{2\, {K_3}\, {\beta_0}\, {\varphi_0}}{\pi} + \left(\frac{5}{18\, \pi} - \frac{5\, \pi}{128}\right)\, {\beta_0}\, {\beta_1}\, {\varphi_0}\right) \nn\\
  & & + \left(a^3\, b - a\, b^3\right)\, \left(\frac{{{\beta_0}}^3\, {{\varphi_0}}^3}{16} + \left(\frac{40}{27\, \pi} - \frac{265\, \pi}{768} + \frac{{\pi}^3}{64}\right)\, {{\beta_0}}^3 - \frac{{\beta_2}\, {\varphi_0}}{16} \right.\nn\\
  & & \qquad \left. + {{\beta_0}}^2\, \left(\frac{52\, {K_2}}{9\, \pi} - \frac{19\, \pi\, {K_2}}{16}\right) + \frac{32\, {K_4}}{3\, \pi} + \left(\frac{2}{9\, \pi} - \frac{\pi}{32}\right)\, {\beta_2} \right.\nn\\
  & & \qquad \left. + {\beta_1}\, \left(\frac{16\, {K_2}}{9\, \pi} - \frac{\pi\, {K_2}}{4}\right) + {\beta_0}\, \left(\frac{32\, {K_3}}{3\, \pi} - \frac{3\, \pi\, {K_3}}{2}\right) - \frac{{K_2}\, {\beta_1}\, {\varphi_0}}{2} - 3\, {K_3}\, {\beta_0}\, {\varphi_0} \right.\nn\\
  & & \qquad \left. - \frac{95\, {\beta_0}\, {\beta_1}\, {\varphi_0}}{192} - \frac{19\, {K_2}\, {{\beta_0}}^2\, {\varphi_0}}{8} + \left(\frac{65}{54\, \pi} - \frac{95\, \pi}{384}\right)\, {\beta_0}\, {\beta_1} - \left(\frac{2}{3\, \pi} - \frac{3\, \pi}{32}\right)\, {{\beta_0}}^3\, {{\varphi_0}}^2 \right.\nn\\
  & & \qquad \left. + \left(\frac{{\pi}^2}{16} - \frac{265}{384}\right)\, {{\beta_0}}^3\, {\varphi_0} - \frac{5\, {\beta_0}\, {\beta_1}\, {{\varphi_0}}^2}{12\, \pi} - \frac{2\, {K_2}\, {{\beta_0}}^2\, {{\varphi_0}}^2}{\pi}\right) \nn\\ 
  & &  + \left(a^5 - 10\, a^3\, b^2 + 5\, a\, b^4\right)\, \left({K_5} + \frac{{{\beta_0}}^4\, {{\varphi_0}}^4}{256} - \frac{3\, {{\beta_1}}^2\, {{\varphi_0}}^2}{512} + {{\beta_0}}^2\, \left(\frac{265\, {K_3}}{192} - \frac{{\pi}^2\, {K_3}}{8}\right) \right.\nn\\
  & & \qquad \left. + \frac{19\, {\beta_3}}{3072} + {{\beta_0}}^3\, \left(\frac{3355\, {K_2}}{4608} - \frac{19\, {\pi}^2\, {K_2}}{192}\right) + \left(\frac{{\pi}^4}{1280} - \frac{265\, {\pi}^2}{9216} + \frac{41041}{221184}\right)\, {{\beta_0}}^4 \right.\nn\\
  & & \qquad \left. - \left(\frac{{\pi}^2}{512} - \frac{265}{12288}\right)\, {{\beta_1}}^2 + \frac{19\, {K_2}\, {\beta_2}}{384} + \frac{19\, {K_3}\, {\beta_1}}{64} + \frac{19\, {K_4}\, {\beta_0}}{12} - \frac{19\, {K_2}\, {{\beta_0}}^3\, {{\varphi_0}}^2}{64} \right.\nn\\
  & & \qquad \left. - \frac{3\, {K_3}\, {{\beta_0}}^2\, {{\varphi_0}}^2}{8} - \frac{247\, {{\beta_0}}^2\, {\beta_1}\, {{\varphi_0}}^2}{3072} + \left(\frac{10}{27\, \pi} - \frac{265\, \pi}{3072} + \frac{{\pi}^3}{256}\right)\, {{\beta_0}}^4\, {\varphi_0} \right.\nn\\
  & & \qquad \left. - \left(\frac{{\pi}^2}{256} - \frac{265}{6144}\right)\, {\beta_0}\, {\beta_2} + \left(\frac{1}{24\, \pi} - \frac{3\, \pi}{512}\right)\, {{\beta_1}}^2\, {\varphi_0} + {{\beta_0}}^3\, {\varphi_0}\, \left(\frac{13\, {K_2}}{9\, \pi} - \frac{19\, \pi\, {K_2}}{64}\right) \right.\nn\\
  & & \qquad \left. + {{\beta_0}}^2\, {\varphi_0}\, \left(\frac{8\, {K_3}}{3\, \pi} - \frac{3\, \pi\, {K_3}}{8}\right) + \left(\frac{{\pi}^2}{128} - \frac{265}{3072}\right)\, {{\beta_0}}^4\, {{\varphi_0}}^2 + \frac{{\beta_3}\, {\varphi_0}}{96\, \pi} - \frac{3\, {\beta_0}\, {\beta_2}\, {{\varphi_0}}^2}{256} \right.\nn\\
  & & \qquad \left. + {\beta_0}\, {\beta_1}\, \left(\frac{1855\, {K_2}}{4608} - \frac{7\, {\pi}^2\, {K_2}}{192}\right) - \left(\frac{247\, {\pi}^2}{9216} - \frac{43615}{221184}\right)\, {{\beta_0}}^2\, {\beta_1} \right.\\
  & & \qquad \left. - \left(\frac{1}{18\, \pi} - \frac{\pi}{128}\right)\, {{\beta_0}}^4\, {{\varphi_0}}^3 + \left(\frac{169}{432\, \pi} - \frac{247\, \pi}{3072}\right)\, {{\beta_0}}^2\, {\beta_1}\, {\varphi_0} - \frac{{K_2}\, {{\beta_0}}^3\, {{\varphi_0}}^3}{6\, \pi} \right.\nn\\
  & & \qquad \left. - \frac{13\, {{\beta_0}}^2\, {\beta_1}\, {{\varphi_0}}^3}{288\, \pi} + \frac{{K_2}\, {\beta_2}\, {\varphi_0}}{12\, \pi} + \frac{{K_3}\, {\beta_1}\, {\varphi_0}}{2\, \pi} + \frac{8\, {K_4}\, {\beta_0}\, {\varphi_0}}{3\, \pi} - \frac{7\, {K_2}\, {\beta_0}\, {\beta_1}\, {{\varphi_0}}^2}{64} \right.\nn\\
  & & \qquad \left. + \left(\frac{1}{12\, \pi} - \frac{3\, \pi}{256}\right)\, {\beta_0}\, {\beta_2}\, {\varphi_0} + {\beta_0}\, {\beta_1}\, {\varphi_0}\, \left(\frac{7\, {K_2}}{9\, \pi} - \frac{7\, \pi\, {K_2}}{64}\right)\right) \nn \\
  & &  + \left(b^5 + 5\, a^4\, b - 10\, a^2\, b^3\right)\, \left(\frac{19\, {{\beta_0}}^4\, {{\varphi_0}}^3}{768} + \left(\frac{121}{324\, \pi} - \frac{3355\, \pi}{36864} + \frac{19\, {\pi}^3}{3072}\right)\, {{\beta_0}}^4 \right.\nn\\
  & &  \qquad \left. + {{\beta_0}}^3\, \left(\frac{40\, {K_2}}{27\, \pi} + \frac{{\pi}^3\, {K_2}}{64} - \frac{265\, \pi\, {K_2}}{768}\right) - \frac{{\beta_3}\, {\varphi_0}}{256} + \left(\frac{13}{288\, \pi} - \frac{19\, \pi}{2048}\right)\, {{\beta_1}}^2 \right. \nn\\
  & & \qquad \left. + {{\beta_0}}^2\, \left(\frac{26\, {K_3}}{9\, \pi} - \frac{19\, \pi\, {K_3}}{32}\right) + \frac{8\, {K_5}}{3\, \pi} - \frac{19\, {{\beta_1}}^2\, {\varphi_0}}{1024} + \left(\frac{1}{72\, \pi} - \frac{\pi}{512}\right)\, {\beta_3} \right.\nn\\
  & & \qquad \left. + {\beta_2}\, \left(\frac{{K_2}}{9\, \pi} - \frac{\pi\, {K_2}}{64}\right) + {\beta_1}\, \left(\frac{2\, {K_3}}{3\, \pi} - \frac{3\, \pi\, {K_3}}{32}\right) + {\beta_0}\, \left(\frac{32\, {K_4}}{9\, \pi} - \frac{\pi\, {K_4}}{2}\right) + \frac{{K_2}\, {{\beta_0}}^3\, {{\varphi_0}}^3}{16} \right. \nn\\
  & & \qquad \left. + \frac{13\, {{\beta_0}}^2\, {\beta_1}\, {{\varphi_0}}^3}{768} + \left(\frac{65}{162\, \pi} - \frac{3445\, \pi}{36864} + \frac{13\, {\pi}^3}{3072}\right)\, {{\beta_0}}^2\, {\beta_1} - \frac{{K_2}\, {\beta_2}\, {\varphi_0}}{32} - \frac{3\, {K_3}\, {\beta_1}\, {\varphi_0}}{16} \right.\nn\\
  & & \qquad \left. - {K_4}\, {\beta_0}\, {\varphi_0} - \frac{19\, {\beta_0}\, {\beta_2}\, {\varphi_0}}{512} - {{\beta_0}}^3\, {\varphi_0}\, \left(\frac{265\, {K_2}}{384} - \frac{{\pi}^2\, {K_2}}{16}\right) + \frac{{{\beta_0}}^4\, {{\varphi_0}}^4}{96\, \pi} -\frac{{{\beta_1}}^2\, {{\varphi_0}}^2}{64\, \pi} \right.\nn\\
  & & \qquad \left. - \frac{19\, {K_3}\, {{\beta_0}}^2\, {\varphi_0}}{16} + \left(\frac{13}{144\, \pi} - \frac{19\, \pi}{1024}\right)\, {\beta_0}\, {\beta_2} + {\beta_0}\, {\beta_1}\, \left(\frac{91\, {K_2}}{108\, \pi} - \frac{133\, \pi\, {K_2}}{768}\right) \right.\nn\\
  & & \qquad \left. - \left(\frac{13}{72\, \pi} - \frac{19\, \pi}{512}\right)\, {{\beta_0}}^4\, {{\varphi_0}}^2 - {{\beta_0}}^3\, {{\varphi_0}}^2\, \left(\frac{2\, {K_2}}{3\, \pi} - \frac{3\, \pi\, {K_2}}{32}\right) \right.\nn\\
  & & \qquad \left. + \left(\frac{19\, {\pi}^2}{768} - \frac{3355}{18432}\right)\, {{\beta_0}}^4\, {\varphi_0} - \frac{{\beta_0}\, {\beta_2}\, {{\varphi_0}}^2}{32\, \pi} - \frac{133\, {K_2}\, {\beta_0}\, {\beta_1}\, {\varphi_0}}{384} - \frac{{K_3}\, {{\beta_0}}^2\, {{\varphi_0}}^2}{\pi} \right.\nn\\
  & & \qquad \left. - \left(\frac{13}{72\, \pi} - \frac{13\, \pi}{512}\right)\, {{\beta_0}}^2\, {\beta_1}\, {{\varphi_0}}^2 + \left(\frac{13\, {\pi}^2}{768} - \frac{3445}{18432}\right)\, {{\beta_0}}^2\, {\beta_1}\, {\varphi_0} - \frac{7\, {K_2}\, {\beta_0}\, {\beta_1}\, {{\varphi_0}}^2}{24\, \pi}\right),\nn
\end{eqnarray}
or numerically:
\begin{eqnarray}
  \label{eq:fopt_mod_num}
  & & \delta_{\rm pert}  =  a + 0.8488\,b + (-4.5\,\varphi_0 + 0.8082)\,a\,b  + \nn\\
  & & (1.9099\,\varphi_0 + 5.2023)\,(a^2 - b^2) + \nn\\
  & & (- 5.0625\,\varphi_0^2 + 5.2138\,\varphi_0 + 26.366)\,(a^3 - 3\,a\,b^2) + \nn\\
  & & (4.2972\,\varphi_0^2 + 27.410\,\varphi_0 + 12.356)\,(b^3 - 3\,a^2\,b) + \\
  & & (- 9.6687\,\varphi_0^3 - 101.51\,\varphi_0^2 - 71.629\,\varphi_0 +
  127.08)\,(a^4 - 6\,a^2\,b^2 + b^4) + \nn\\
  & & (45.563\,\varphi_0^3 - 100.94\,\varphi_0^2 - 918.59\,\varphi_0 - 521.11)\,(a^3\,b - a\,b^3) + \nn\\
  & & (25.629\,\varphi_0^4 - 92.897\,\varphi_0^3 - 1220.5\,\varphi_0^2 - 1272.5\,\varphi_0 + \nn\\
  & & \qquad K_5 + 307.78)\,(a^5 - 10\,a^3\,b^2 + 5\,a\,b^4) + \nn\\
  & & (21.755\,\varphi_0^4 + 324.78\,\varphi_0^3 + 271.83\,\varphi_0^2 - 1612.0\,\varphi_0 + \nn\\ 
  & & \qquad 0.8488\,K_5 - 1413.5)\,(b^5 + 5\,a^4\,b - 10\,a^2\,b^3),\nn
\end{eqnarray}
with $\varphi_0\in[-\pi,0]$, and $\alphas(m_\tau^2\exp(i
\varphi_0))/\pi = a + ib$.  Three points should be noted about
equation~(\ref{eq:fopt_mod}):
\begin{enumerate}
\item it resembles the usual FOPT result for $\varphi_0=0$ and $b=0$;
\item $\delta_{\rm pert}$ remains real for all choices of $\alphas$
  and $\varphi_0$;
\item inserting the Taylor expanded $\alphas(\varphi_0)$ in
  eq.~(\ref{eq:fopt_mod}) and Taylor expanding the resulting
  $\delta_{\rm pert}$ again around $\alphas(\varphi_0=0)$ leads also
  to the usual FOPT result.
\end{enumerate}
The last point demonstrates that FOPT can be generalized only if the
\lq exact\rq\ value for $\alphas(\varphi_0)$ is used in the expansion.

\begin{figure}[htb]
  \begin{center}
    \resizebox{0.8\textwidth}{!}{%
      \includegraphics{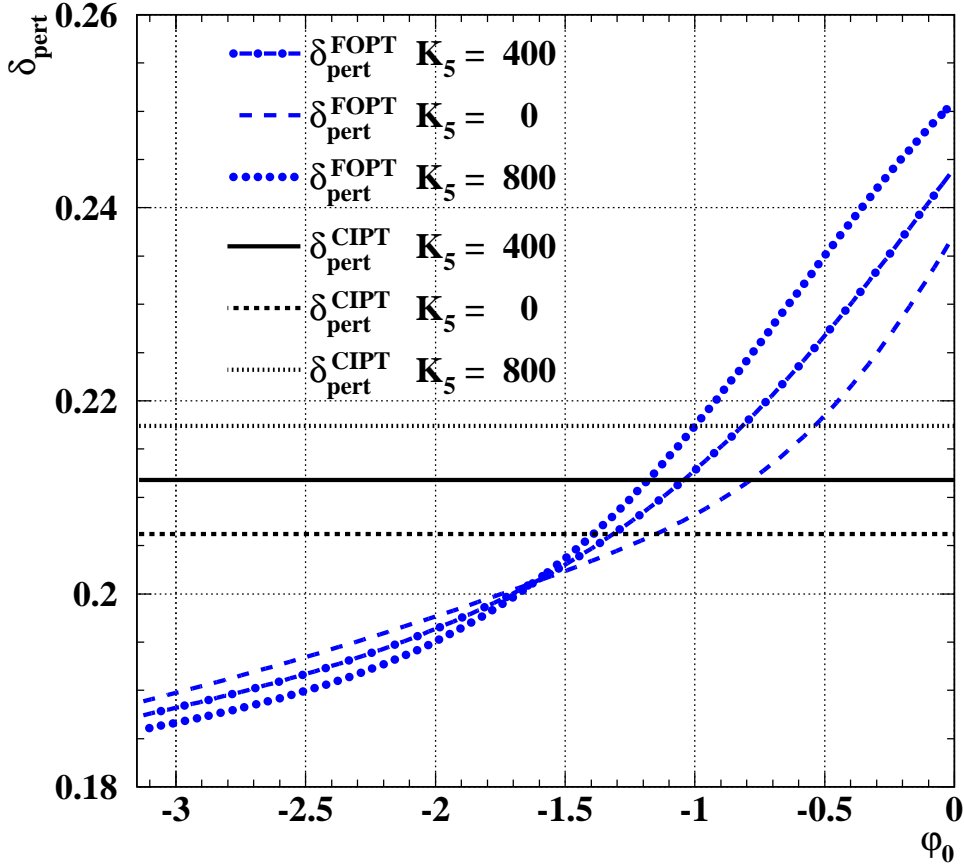}}
    \caption{\it $\delta_{\mathrm{pert}}$ as function of the
      development point $\varphi_0$. The solid, medium-dashed and
      short-dashed lines show the CIPT result to 5th order for $K_5 =
      400$, $0$, and $800$, respectively.  The long-dashed,
      dash-dotted and dotted lines show the generalized FOPT result to
      5th order for $K_5 = 400$, $0$, and $800$, respectively.  The
      reference value $\alphas(m_\tau^2) = 0.35$ was used for all
      curves.}
    \label{fig:pert}
  \end{center}
\end{figure}
Figure~\ref{fig:pert} shows $\delta_{\rm pert}$ as a function of
$\varphi_0$ with $\alphas(m_\tau^2)=0.35$ and $K_5 = 400,0,800$ as
reference values. The consequences of the generalized FOPT solution
are discussed in the following section.

\section{Discussion of the generalized FOPT solution}
As can be seen from figure~\ref{fig:pert} the FOPT result depends
largely on the choice of $\varphi_0$. The FOPT curves intersect with
the CIPT curves around $\varphi_0\simeq-1$ but span over a much larger
range of $\delta_{\rm pert}$ values. Compared to the uncertainty from
the neglected higher orders this intrinsic error is $4$ times larger
as none of the choices for $\varphi_0$ should be excluded.  The
default choice of $\varphi_0 = 0$ leads to the largest possible value
of $\delta_{\rm pert}$ and therefore \alphas\ from FOPT used to be
smaller than from CIPT. The deviation can however not be attributed to
higher order terms in the series of $\delta_{\rm pert}$.  Instead the
extraction of \alphas\ with FOPT should use the average of the two
extremes $\alphas(\varphi_0=0)$ and $\alphas(\varphi_0=-\pi)$ and half
of their difference as additional theoretical error.  Consequently the
most accurate way for the determination of \alphas\ from $\tau$ decays
is the CIPT approach. There is no reason to add the same error to the
CIPT result as it does not depend on the choice of $\varphi_0$. Also
since the FOPT result agrees within its own error with CIPT there is
no discrepancy anymore between results with these two approaches.  In
fact the CIPT result is what FOPT would converge to for $n\to\infty$
if $n$ equidistant points on the circle $s=s_0\exp(i\varphi)$ would be
used in the expansion. The case $n=1$ could therefore be regarded as
an approximation for CIPT and the choice $\varphi_0 = 0$ is just one
of the many possible choices for $n=1$.

\section{Renormalon Chains}
The third theory often used in evaluating $\alphas$ from $\tau$ decays
uses so called \lq Renormalon
Chains\rq~\cite{Ball:1995ni,Neubert:1996gd,Lovett-Turner:1995ti,Maxwell:1996ig}
and re-sums the $\beta_0$ parts of $\delta_{\rm pert}$ to all orders
in $\alphas$. In~\cite{Neubert:1996gd} the result obtained from this
re-summation is corrected by the known FOPT terms by first subtracting
the large-$\beta_0$ part up to the desired order of FOPT and then
adding the FOPT part. Thus $\delta_{\rm pert}$ for the Renormalon
Chain Perturbation Theory (RCPT) can be written as
\begin{equation}\label{eq:dpert_RCPT}
  \delta_{\rm pert}^{\rm RCPT} = \delta_{\rm renormalon} 
  - \delta_{{\rm large-}\beta_0}^{\rm FOPT} 
  + \delta_{\rm pert}^{\rm FOPT},
\end{equation}
where the three terms in the sum refer to the renormalon chain result,
the large-$\beta_0$ re-summed result up to the order used in FOPT, and
the FOPT result, respectively.  As is pointed out
in~\cite{Neubert:1996gd} the renormalon chain re-summation includes
parts of the CIPT re-summation, namely the terms $(-\beta_0/4
\ln(s/s_0))^n$ which are part of the coefficient in front of
$(\alphas/\pi)^n$ in eq.~(\ref{eq:a(s)_taylor}). Therefore the CIPT
result can not be used instead of the FOPT result in
eq.(\ref{eq:dpert_RCPT}). Still, for the FOPT correction and the fixed
order large-$\beta_0$ correction the same arbitrariness of the choice
of $\varphi_0$ as discussed in the first part of this note exists, as
long as both the FOPT term and the fixed order large-$\beta_0$ term
are expanded around the same $\varphi_0$.  Therefore the variation of
$\delta_{\rm pert}^{\rm FOPT}- \delta_{{\rm large-}\beta_0}^{\rm
  FOPT}$ with $\varphi_0$ is a source of uncertainty in the RCPT
approach.  The generalized $\delta_{{\rm large-}\beta_0}^{\rm FOPT}$
can be derived from eq.~(\ref{eq:fopt_mod}) by setting $\beta_n=0$ for
$n>0$ and replacing the $K_n$ with $\beta_0^{(n-1)}\kappa_n$, which
are given up to $n=4$ in~\cite{Neubert:1996gd} and up to $n=12$
in~\cite{Beneke:2008ad}. Numerically $\delta_{{\rm
    large-}\beta_0}^{\rm FOPT}$ up to the fifth order in $\alphas$ is
given by:
\begin{eqnarray} \label{eq:rcpt_mod_num}
  & &      \delta_{\beta_0}^{\rm FOPT}  =  a + 0.8488\,b + (- 4.5\,\varphi_0 + 0.6668)\,a\,b + \nn\\
  & &  (1.9099\,\varphi_0 + 5.1190)\,(a^2 - b^2) + \nn\\
  & &  (- 5.0625\,\varphi_0^2 + 1.5002\,\varphi_0 + 28.779)\,(a^3 - 3\,a\,b^2) + \nn\\
  & &  (4.2972\,\varphi_0^2 + 23.035\,\varphi_0 + 2.5067)\,(b^3 - 3\,a^2\,b) + \\
  & & (- 9.6687\,\varphi_0^3 - 77.745\,\varphi_0^2 - 16.920\,\varphi_0
  + 
  156.67)\,(a^4 - 6\,a^2\,b^2 + b^4) + \nn\\
  & &  (45.563\,\varphi_0^3 - 20.253\,\varphi_0^2 - 777.03\,\varphi_0 - 433.69)\,(a^3\,b - a\,b^3) + \nn\\ 
  & &  (25.629\,\varphi_0^4 - 15.189\,\varphi_0^3 - 874.16\,\varphi_0^2 - 975.80\,\varphi_0 + \nn\\
  & & \qquad 900.78)\,(a^5 - 10\,a^3\,b^2 + 5\,a\,b^4) + \nn\\
  & &  (21.755\,\varphi_0^4 + 233.23\,\varphi_0^3 + 76.141\,\varphi_0^2 - 1410.1\,\varphi_0 - \nn\\
  & & \qquad 615.93)\,(b^5 + 5\,a^4\,b - 10\,a^2\,b^3),\nn 
\end{eqnarray}
with $\varphi_0\in[-\pi,0]$, and $\alphas(m_\tau^2\exp(i
\varphi_0))/\pi = a + ib$ as in eq.~(\ref{eq:fopt_mod_num}).  The
difference $\delta_{\rm pert}^{\rm FOPT}-\delta_{{\rm large-}\beta_0}$
is not as sensitive to the choice of $\varphi_0$ as $\delta_{\rm
  pert}^{\rm FOPT}$ alone and roughly halfs the associated uncertainty
in \alphas.

\begin{figure}[htb]
  \begin{center}
    \resizebox{0.8\textwidth}{!}{%
      \includegraphics{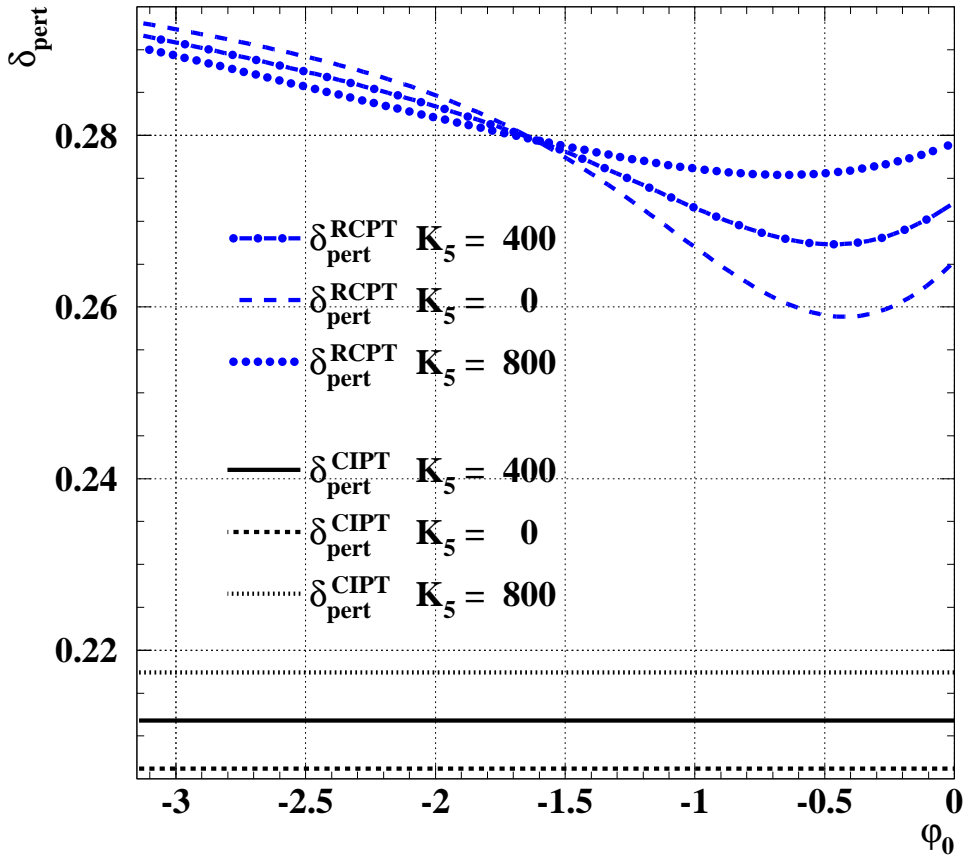}}
    \caption{\it $\delta_{\mathrm{pert}}$ as function of the
      development point $\varphi_0$. The solid, medium-dashed and
      short-dashed lines show the CIPT result to 5th order for $K_5 =
      400$, $0$, and $800$, respectively.  The long-dashed,
      dash-dotted and dotted lines show the RCPT result with 0-loop
      matching and generalized FOPT correction to 5th order for $K_5 =
      400$, $0$, and $800$, respectively.  The reference value
      $\alphas(m_\tau^2) = 0.35$ was used for all curves.}
    \label{fig:rcpt_1}
  \end{center}
\end{figure}
Figure~\ref{fig:rcpt_1} shows the RCPT result using the $\delta_{\rm
  renormalon}$ as in~\cite{Neubert:1996gd} but the modified
$\delta_{\rm pert}^{\rm FOPT}-\delta_{{\rm large-}\beta_0}$ from
eqs.~(\ref{eq:fopt_mod_num},\ref{eq:rcpt_mod_num}) to correct the
result up to the fifth order in $\alphas$.  The reference value of
$\alpha_s(m_\tau^2) = 0.35$ is used again for all curves.  It is clear
from the figure that RCPT would still require a much smaller
$\alphas\simeq0.31$ compared to CIPT even with the modified
corrections in the fixed order parts.

This observation relies however on the fact that $\alphas$ in the
renormalon part and the fixed order part of eq.~(\ref{eq:dpert_RCPT})
refers to the same quantity. This is probably not the case. The
renormalon part in~\cite{Neubert:1996gd} is derived from the one-loop
coupling in the so-called V scheme, $\alpha_{\rm s}^{\rm V}(\mu^2)$
which is matched on the 0-loop level to $\alpha_{\rm s}^{\overline{\rm
    MS}}(\exp(-5/3)\mu^2)$. The problem therefore is that we have a
coupling constant on the 3-loop level\footnote{Since $\beta_3$ enters
  only in the 5th order in $R_\tau$ $\alphas$ is effectively used as a
  3-loop coupling constant in FOPT which goes up to the 4th (plus
  estimated 5th) order in \alphas} in the FOPT parts, but treat it as
a one-loop coupling in the renormalon parts.  A possible solution
would be to use 2-loop matching to go from the $\overline{\rm
  MS}$-scheme to the V scheme which is given
by~\cite{Peter:1997me,Schroder:1998vy}
\begin{equation}\label{eq:as_V}
  \frac{\D\alpha_{\rm s}^{\rm V}\left(\mu^2\right)}{\D\pi} = \frac{\D\alpha_{\rm s}^{\overline{\rm MS}}\left(e^{-5/3} \mu^2\right)}{\D\pi} - 2\left(\frac{\alpha_{\rm s}^{\overline{\rm MS}}\left(e^{0.4221}\mu^2\right)}{\D\pi}\right)^2 - 7.72816\left(\frac{\alpha_{\rm s}^{\overline{\rm MS}}\left(e^{0.4221}\mu^2\right)}{\D\pi}\right)^3.
\end{equation}

\begin{figure}[htb]
  \begin{center}
    \resizebox{0.8\textwidth}{!}{%
      \includegraphics{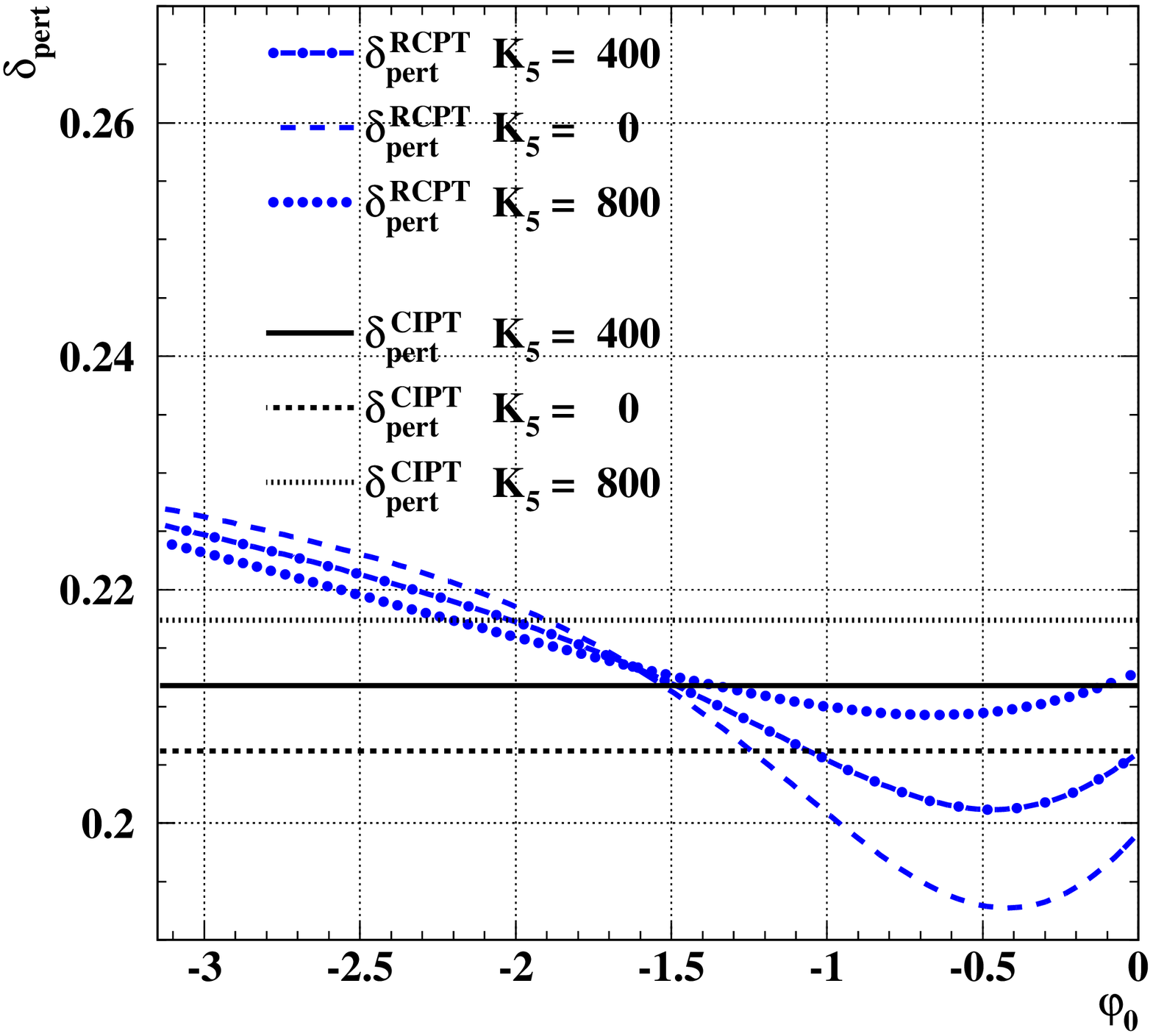}}
    \caption{\it $\delta_{\mathrm{pert}}$ as function of the
      development point $\varphi_0$. The solid, medium-dashed and
      short-dashed lines show the CIPT result to 5th order for $K_5 =
      400$, $0$, and $800$, respectively.  The long-dashed,
      dash-dotted and dotted lines show the RCPT result with 2-loop
      matching and generalized FOPT correction to 5th order for $K_5 =
      400$, $0$, and $800$, respectively.  The reference value
      $\alphas(m_\tau^2) = 0.35$ was used for all curves.}
    \label{fig:rcpt_2}
  \end{center}
\end{figure}
Figure~\ref{fig:rcpt_2} shows again the RCPT result as before but with
the 2-loop matching for $\alpha_{\rm s}^{\rm V}$.  The reference value
of $\alpha_s(m_\tau^2) = 0.35$ is used for all RCPT and CIPT curves.
The large overlap of the CIPT and RCPT curves shows that the
differences in the deduced strong couplings from both theories are
much smaller than previously assumed.

\section{Discussion of the modified RCPT solution}
\begin{figure}[htb]
  \begin{center}
    \resizebox{0.8\textwidth}{!}{%
      \includegraphics{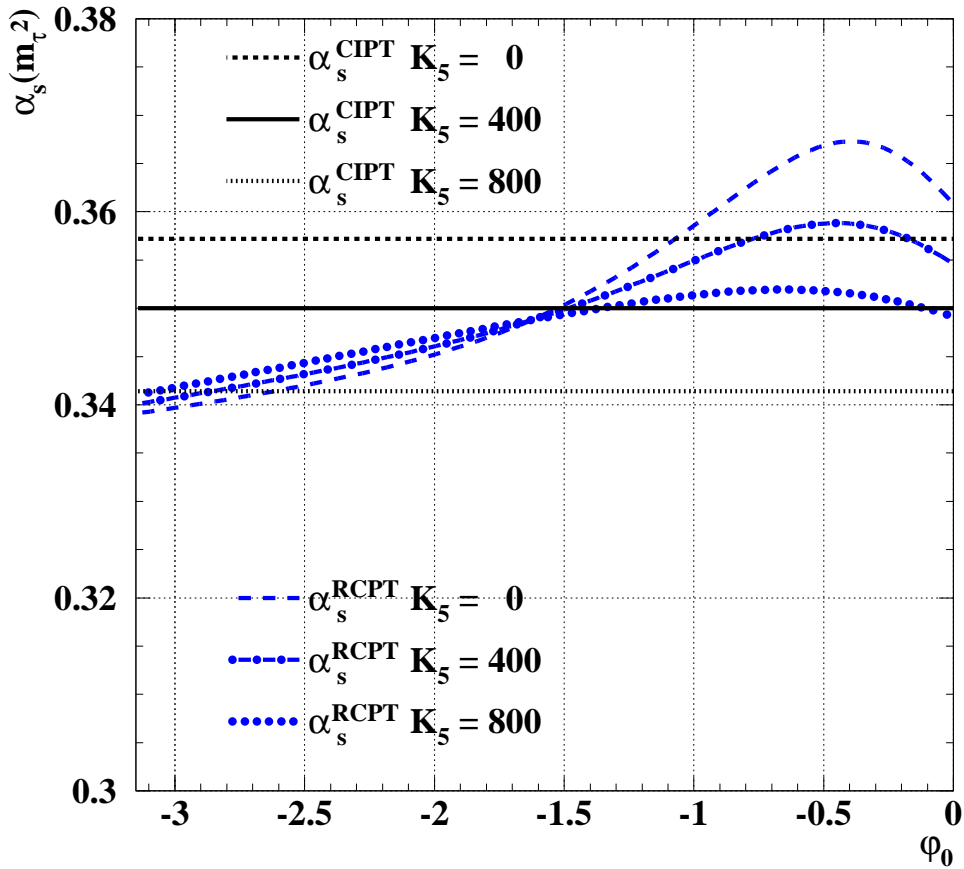}}
    \caption{\it $\alphas(m_\tau^2)$ as function of the development
      point $\varphi_0$. The solid, medium-dashed and short-dashed
      lines show the CIPT result to 5th order for $K_4 = 400$, $0$,
      and $800$, respectively.  The long-dashed, dash-dotted and
      dotted lines show the RCPT result with 2-loop matching and
      generalized FOPT correction to 5th order for $K_5 = 400$, $0$,
      and $800$, respectively.  All curves are obtained with
      $\delta_{\rm pert} = 0.21179$ as reference value.}
    \label{fig:alphas_fit}
  \end{center}
\end{figure}
Unlike in the case of the generalized FOPT the RCPT solution can not
be regarded as the first iteration of a contour improved result as
this would result in inconsistent definitions of $\alphas$ on the
circle $s=|m_\tau^2|$. Therefore the magnitude of the spread of
$\alphas$ values obtained from a fixed $\delta_{\rm pert}$ and a fixed
$K_5$ could be regarded as induced by higher order terms. This spread
has furthermore the same magnitude as the difference between CIPT and
RCPT at $\varphi_0=0$, which at the same time shrunk by a factor of
six by using the 2-loop matching instead of 0-loop matching for the
transition from the V scheme to the $\overline{\rm MS}$ scheme.
Averaging over all $\varphi_0=0$ values for RCPT leads to the same
numerical value for $\alphas$ as CIPT at the central choice for $K_5$,
while the RCPT results for $K_5 = 0$, $800$ stay much closer to the
central \alphas\ compared to the corresponding CIPT values, showing
that the large $\beta_0$ re-summation reduces the impact of higher
order terms.

Recently the situation about the influence of higher-order corrections
has been revisited in~\cite{Beneke:2008ad}, where the authors study
the renormalon structure in different models and fix the Borel
transform of the Adler function such that the known terms up to 4th
order are reproduced.  Using this matched Adler function for $R_\tau$
and comparing the full re-summed result with standard FOPT shows again
good agreement, while the distance to CIPT is large. But as discussed
in section~\ref{sec:genFOPT}, the generalized FOPT solution would on
average reproduce the CIPT result and therefore the model
of~\cite{Beneke:2008ad} would also deviate from the approach advocated
in section~\ref{sec:genFOPT}.

The power corrections to $R_\tau$ have been re-examined
in~\cite{Narison:2009vy} where duality violation
parts~\cite{Cata:2008ye,Cata:2008ru} of the order of $0.01$ (but found
to be negligible in~\cite{Davier:2008sk}) and tachyonic mass
corrections from the gluon are considered. It is argued
in~\cite{Narison:2009vy} that the difference between the Borel-sum and
the truncated series at 4th order in the large-$\beta_0$ limit can be
regarded as non-standard dimension 2 power corrections to
$R_\tau$. Since these corrections are of the order $0.04$ for CIPT and
$0.02$ for FOPT, respectively, they dominate over the duality
violation effects and if taken at face value eliminate the difference
between FOPT and CIPT. However, it should be noted that the Borel
transform $\hat{D}(b)$ of the Adler function does not have a pole at
$b=1$ and therefore no renormalon ambiguity of dimension 2, which
makes it difficult to associate a dimension 2 correction to the
observed difference.

Here I'll concentrate on the perturbative parts only and neglect any
non-standard power correction. Figures~\ref{fig:pert}
and~\ref{fig:rcpt_2} show that there is no large cancellation
mechanism which would prevent the FOPT or the RCPT result from
depending strongly on the arbitrary choice of the development point
$\varphi_0$. Including this arbitrary choice in the uncertainty
estimate shows that CIPT provides the most accurate estimate. It is
however re-assuring that correcting the large-$\beta_0$ re-summed
result with the known fixed order terms up to 5th order reduces this
dependency by $50\%$ leaving the remaining $50\%$ to the exact
logarithms for $\beta_n$ with $n>0$.

Figure~\ref{fig:alphas_fit} shows a numerical example for $\delta_{\rm
  pert}=0.21179$ comparing the values of $\alphas$ for $K_5=400,0,800$
from fits to CIPT and RCPT as a function of $\varphi_0$.  The CIPT
numbers are:
\begin{eqnarray}\label{eq:as_fit_CIPT}
  \alphas(m_\tau^2,\delta_{\rm pert}= 0.21179, K_5=\phantom{20}0)^{\rm CIPT} & = & 0.3572, \nn \\
  \alphas(m_\tau^2,\delta_{\rm pert}= 0.21179, K_5=400)^{\rm CIPT} & = & 0.35, \\
  \alphas(m_\tau^2,\delta_{\rm pert}= 0.21179, K_5=800)^{\rm CIPT} & = & 0.3432, \nn 
\end{eqnarray}
and for RCPT the result is:
\begin{eqnarray}\label{eq:as_fit_RCPT}
  \alphas(m_\tau^2,\delta_{\rm pert}= 0.21179, K_5=\phantom{20}0)^{\rm RCPT} & = & 0.3519\pm0.0097, \nn \\
  \alphas(m_\tau^2,\delta_{\rm pert}= 0.21179, K_5=400)^{\rm RCPT} & = & 0.3498\pm0.0063, \\
  \alphas(m_\tau^2,\delta_{\rm pert}= 0.21179, K_5=800)^{\rm RCPT} & = & 0.3480\pm0.0034, \nn 
\end{eqnarray}
where the errors are given by the RMS of the $\alphas$ values over the
range $-\pi\le\varphi_0\le0$.  Now RCPT and CIPT agree on the central
value of \alphas\ and both give similar estimates for the uncertainty
due to (different) neglected higher order terms:
\begin{equation}\label{eq:as_final}
  \alphas(m_\tau^2,\delta_{\rm pert}= 0.21179) = 0.3499\pm0.0072{{}^{+0.0052}_{-0.0007}}_\mu.
\end{equation}
where the first error is due to $K_5\simeq400\pm400$ and the second
due to the variation of the renormalization scale $0.4\le
\mu^2/m_\tau^2\le 1.6$.
\section{Numerical analysis}
Using the same numerical value for $\delta_{\rm
  pert}=0.2042\pm0.0038_{\rm exp}\pm0.0033_{\rm non-pert}$ as obtained
in~\cite{Davier:2008sk} and used in~\cite{Beneke:2008ad}, where the
first error is the experimental one, dominated by the non-strange
hadronic decay ratio of the $\tau$, $R_{\tau,V+A}$ and the second is
due to the non-perturbative and quark-mass corrections, the results
for CIPT and generalized FOPT and RCPT read:
\begin{eqnarray}
  \alpha_{\rm s}^{\rm CIPT}(m_\tau^2) & = & 0.3406 \pm 0.0047_{\rm exp} \pm 0.0041_{\rm non-pert} \pm 0.0066_{K_5}, \nn\\
  \alpha_{\rm s}^{\rm FOPT}(m_\tau^2) & = & 0.3535 \pm 0.0061_{\rm exp} \pm 0.0053_{\rm non-pert} \pm 0.0208_{\varphi_0}{{}^{+0.0005}_{-0.0001}}_{K_5},\\
  \alpha_{\rm s}^{\rm RCPT}(m_\tau^2) & = & 0.3440 \pm 0.0030_{\rm exp} \pm 0.0026_{\rm non-pert} \pm 0.0061_{\varphi_0} \pm 0.0019_{K_5}. \nn
  ,\label{eq:alphas_num}
\end{eqnarray}

All three results agree within the error due to $\varphi_0$ which is
very large for FOPT but moderate in case of RCPT. The difference
between CIPT and RCPT is of the same size as the error due to
$\varphi_0$ for RCPT and the average between both values (and
conservatively assigning the larger of the two results errors to the
average) leads to:
\begin{equation}\label{eq:as_fit_num}
  \alphas(m_\tau^2)  =  0.3423 \pm 0.005_{\rm exp} \pm 0.007_{\Delta K_5} \pm 0.004_{\rm non-pert}{{}^{+0.005}_{-0.001}}_\mu,
\end{equation}
where the fourth error is due to the variation of the renormalization
scale. The total theoretical error (including the non-perturbative
part) is with $\pm0.008$ only marginally larger than the experimental
error.

Evolving $\alphas$ given by eq.~(\ref{eq:as_fit_num}) from
$m_\tau=1.7768\,{\rm GeV}$ to $m_{\rm Z^0}=91.1876$ with $m_{\rm
  c}(m_{\rm c}) = 1.27^{+0.07}_{-0.11}\,{\rm GeV}$ and $m_{\rm
  b}(m_{\rm b}) = 4.20^{+0.17}_{-0.07}\,{\rm
  GeV}$~\cite{Amsler:2008zzb} with the flavor thresholds to $n_f=4$
and $n_f=5$ at $m_\tau$ and $m_{\rm b}$, respectively, gives:
\begin{equation}\label{eq:as_fit_num_Z}
  \alphas(m_{\rm Z^0}^2)  =  0.1213 \pm 0.0006_{\rm exp} \pm 0.0008_{\Delta K_5} \pm 0.0004_{\rm non-pert}{{}^{+0.0005}_{-0.0001}}_\mu \pm 0.0002_{\rm ev},
\end{equation}
where the last error is the evolution uncertainty due to the variation
of the thresholds $m_q < m_{\rm thresh} < 2m_q$ and the quark masses
itself within their respective errors.

\section{Conclusions}
Modifying the usual Taylor expansion in fixed order perturbation
theory by allowing starting points other than $\alphas(m_\tau^2)$ on
the complex circle $|s|=m_\tau^2$ reveals a larger intrinsic
uncertainty of FOPT than previously assumed. Giving equal weight to
all possible choices and averaging over the different values of
$\alphas^{\rm FOPT}$ so obtained brings the FOPT result in agreement
with CIPT. Since CIPT does not bear this additional intrinsic
uncertainty the CIPT solution should be preferred.  The
large-$\beta_0$ re-summed result can be modified in its fixed order
parts with a similar approach. Here the variation due to the starting
point of the strong coupling on the complex circle alone does no
account for the difference to CIPT. But applying 2-loop matching
instead of 0-loop matching to combine the large-$\beta_0$ parts with
the known fixed order parts up to $\beta_2$ cancels the difference of
RCPT and CIPT.  In this way all three perturbative approaches finally
agree on the central value of $\alphas$ from the $\tau$. The final
result from eq.~(\ref{eq:as_fit_num_Z}) is
\begin{equation}\label{eq:as_fit_num_Z2}
  \alphas(m_{\rm Z^0}^2)  =  0.1213 \pm 0.0006_{\rm exp} \pm 0.0010_{\rm theo},
\end{equation}
where the first error is experimental and the second theoretical
including power corrections. This confirms with different theoretical
arguments the large $\alphas$ obtained in~\cite{Davier:2008sk} and is
not compatible with the numerically lower values
in~\cite{Narison:2009vy,Beneke:2008ad,Maltman:2008nf,Maltman:2008ud}.
\section*{Acknowledgements}
I'd like to thank Teresa Barillari, Siggi Bethke and Stefan Kluth for
reading the note and providing useful comments. Many thanks also to
Matthias Jamin for finding typos in
equations~(\ref{eq:kn},\ref{eq:a(s)_taylor}) and to Matthias Jamin and
Martin Beneke for providing corrections for my comments
on~\cite{Beneke:2008ad}.

\bibliographystyle{phppnp} \bibliography{fopt}

\end{document}